\documentclass[aps,prl,twocolumn,groupedaddress,showpacs,notitlepage]{revtex4-1} % remove notitlepage to move toc to a separate page than the title
\usepackage{graphicx}
\usepackage{subfigure}
\usepackage{amsmath}
\usepackage{braket}
\usepackage[colorlinks,linkcolor={blue},citecolor={blue},urlcolor={blue}]{hyperref}

\usepackage{array}
\usepackage{MnSymbol}

\usepackage{bbm} % this is for the command \mathbbm{}
\usepackage{chngcntr} % this is for \counterwithin command in the appendix
\usepackage[page,toc,titletoc,title]{appendix}

\usepackage{array}
\newcolumntype{M}[1]{>{\centering\arraybackslash}m{#1}}

\usepackage[table]{xcolor} % this is for adding background color to table
\usepackage{makecell}

\usepackage{ulem} % to strikethrough text

\newcommand{\bA}{\mathbf{A}}

\newcommand{\br}{\mathbf{r}}

\newcommand{\bp}{\mathbf{p}}

\newcommand{\bZ}{\mathbf{Z}}

\newcommand{\bsigma}{\boldsymbol{\sigma}}
\newcommand{\bmu}{\boldsymbol{\mu}}

\newcommand{\bpi}{\boldsymbol{\pi}}
\newcommand{\bPi}{\boldsymbol{\Pi}}
\newcommand{\bCalA}{\boldsymbol{\mathcal{A}}}
\newcommand{\bCalB}{\boldsymbol{\mathcal{B}}}
\newcommand{\bCalV}{\boldsymbol{\mathcal{V}}}
\newcommand{\bCalK}{\boldsymbol{\mathcal{K}}}

\newcommand{\bl}{\mathbf{l}}
\newcommand{\bv}{\mathbf{v}}

\begin{document}

%\title{Non-Abelian Berry phase effects in inhomogeneously strained moir\'e pattern}
\title{Layer pseudospin dynamics and genuine non-Abelian Berry phase in inhomogeneously strained moir\'e pattern}

\author{Dawei Zhai}
\email{dzhai@hku.hk}

\author{Wang Yao}
\email{wangyao@hku.hk}
\affiliation{Department of Physics, The University of Hong Kong, and HKU-UCAS Joint Institute of Theoretical and Computational Physics at Hong Kong, China}

\date{\today}

\begin{abstract}
Periodicity of long wavelength moir\'e patterns is very often destroyed by the inhomogeneous strain introduced in fabrications of van der Waals layered structures. We present a framework to describe massive Dirac fermions in such distorted moir\'e pattern of transition metal dichalcogenides homobilayers, accounting for the dynamics of layer pseudospin. In decoupled bilayers, we show two causes of in-plane layer pseudospin precession: By the coupling of layer antisymmetric strain to valley magnetic moment; and by the Aharonov-Bohm effect in the SU(2) gauge potential for the case of R-type bilayer under antisymmetric strain and H-type under symmetric strain. With interlayer coupling in the moir\'e, its interplay with strain manifests as a non-Abelian gauge field. We show a genuine non-Abelian Aharonov-Bohm effect in such field, where the evolution operators for different loops are non-commutative. 
This provides an exciting platform to explore non-Abelian gauge field effects on electron, with remarkable tunability of the field by strain and interlayer bias. 
\end{abstract}

%\pacs{}
\maketitle

\begin{figure}[t]
	\includegraphics[width=3.4in]{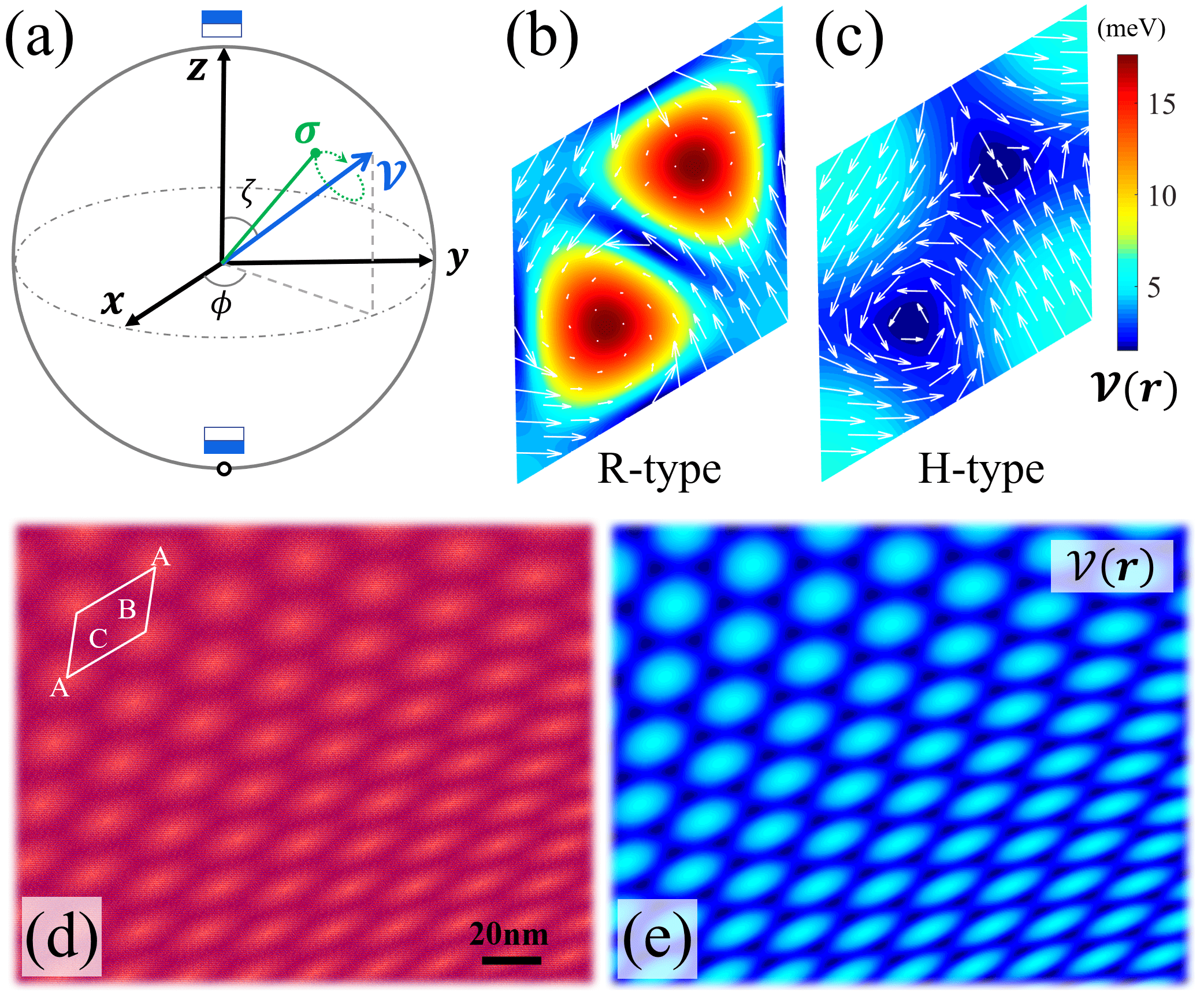}
	\caption{Effective moir\'e field and strained moir\'e pattern. (a) Schematics of layer pseudospin $\bsigma$ (green line) precessing around the moir\'e field $\bCalV$ (blue arrow). (b) Color map plots $\mathcal{V}$ in a rigid moir\'e unit cell, for valley electrons in R-type twisted bilayer MoSe$_2$. Arrows denote the in-plane components of $\hat{\bCalV}$. (c) Same plot for H-type MoS$_2$. (d) An example of strained moir\'e pattern (adapted from Ref.~\cite{YusongStrainInTMDmoire}). The parallelogram denotes a distorted moir\'e cell. (e) The moir\'e field of (c) mapped onto such distorted moir\'e pattern.}
	\label{Fig:StrainedMoireAndMoirePotential}
\end{figure}

Long-wavelength moir\'e patterns by van der Waals stacking of graphene and transition metal dichalcogenides (TMDs) 
have led to the observation of a plethora of novel electron correlation phenomena~\cite{CaoYuanSuperconductivity,CaoYuanCorrelated,FerromagnetismTwistedBilayerGrapheneScience,CoryDeanSuperconductivityScience,AHETwistedBilayerGraphene,MagnetismTwistedBilayerGrapheneNature,HubbardModelMakNature2020,MottWignerCrystalNature2020,hBNspacer}, 
as well as moir\'e excitons as highly tunable quantum emitters~\cite{HongyiMoireExcitonSciAdv2017,XiaodongXuMoireExcitonNature2019,XiaoqinLiMoireExcitonNature2019,FengWangMoireExcitonNature2019,FalkoMoireExcitonNature2019,PhilipKimInterlayerExcitonScience2019}.
In these findings, the moir\'e pattern is exploited as a superlattice energy landscape to trap electrons and excitons, arising from the spatially modulated interlayer coupling~\cite{InterlayerCouplingTwistedTMDSciAdv2017}. 
Theoretical studies revealed that interlayer coupling in the moir\'e manifests as a location dependent Zeeman field on the layer pseudospin (e.g. Fig.~\ref{Fig:StrainedMoireAndMoirePotential}), which exhibits a skyrmion texture in real space~\cite{WuMacDonaldPRL2019,HongyiPseudoFieldMoire,ZhaiPRMaterials}.
For holes, Berry curvature from the adiabatic motion in such moir\'e field is an Abelian gauge field that realizes fluxed superlattices~\cite{HongyiPseudoFieldMoire,ZhaiPRMaterials}, underlying the quantum spin Hall effect discovered in low energy mini-bands~\cite{WuMacDonaldPRL2019,LiangFuStrainMoire}. Link between moir\'e induced gauge field on massless Dirac fermions and flattening of mini-bands in twisted bilayer graphene have also been explored~\cite{PseudFieldBilayerGrapheneLado2019,PseudoFieldTwistedBilayerDaiXi,NonAbelianGrapheneBilayerPRL2012}. For electrons in homobilayer TMDs, the moir\'e field has similar spatial texture but is much weaker at certain spots in the supercell [Figs.~\ref{Fig:StrainedMoireAndMoirePotential}(b--c)]~\cite{InterlayerCouplingTMDPRB2017}, where non-adiabatic pseudospin dynamics need to be accounted. This points to the relevance of intriguing SU(2) Berry phases on massive particles in the moir\'e.
In particular, genuine non-Abelian gauge field-- origin of the noncommutativity of successive loop operations-- in real space is of great interest~\cite{GoldmanNonAbelianReview}, but has only been realized recently in synthetic optical systems~\cite{NonAbelianABEffectScience2019}.  

In experimental reality, inhomogeneous strain is often unintentionally introduced~\cite{BubblesWrinklesInTwistedBilayersNanoLett2014,YusongStrainInTMDmoire}. In monolayers, strain can be described by a pseudo-vector potential with valley contrasted signs, which is associated with an effective magnetic field when strain is inhomogeneous~\cite{StrainPhysRep2016}.
In the moir\'e, dramatic distortion of the periodic landscape occurs in the presence of layer dependent heterostrain~\cite{QingjunTongNatPhys2017}. 
Fig.~\ref{Fig:StrainedMoireAndMoirePotential}(d) is an example of a $0.86^\circ$ twisted moir\'e subject to inhomogeneous heterostrain of peak magnitude $\sim 0.5\%$, where the moir\'e wavelength varies appreciably over a few periods.
For carrier dynamics in such moir\'e, the momentum space description in terms of mini-bands is not validated with the broken periodicity. 

Here we present a framework to describe massive Dirac fermions with the non-adiabatic layer pseudospin dynamics in inhomogeneously distorted TMD homobilayer moir\'e. We first outline several Abelian SU(2) Berry phase effects purely from the layer symmetric and antisymmetric components of strain under R-type (parallel) and H-type (antiparallel) stacking. 
Coupling of the electron's valley magnetic moment to the strain induced magnetic field causes in-plane precession of the pseudospin in layer antisymmetric strain. 
Moreover, R-type (H-type) bilayer under antisymmetric (symmetric) strain features a SU(2) gauge potential in which the Aharonov-Bohm (AB) interference also manifests as in-plane pseudospin precession. When interlayer coupling is considered, its interplay with the strain can be formulated in terms of a non-Abelian gauge field, whose matrix forms at different locations are noncommutative. Evolution in this field has the genuine non-Abelian AB effect where the evolution operators for different loops are noncommutative~\cite{GoldmanNonAbelianReview}. %This distinguishes the strain distorted bilayer moir\'e from decoupled bilayers, and from unstrained moir\'e that features non-Abelian gauge potential with zero gauge field~\cite{ZhaiPRMaterials}.

%%%%%%%%%%%%%%%%%%%%%%%%%%%%%%%%%%%%%%%%%%%%%%%%%%%%%%%%%%%%%%%%%%%%%%%%%%%%%%%%%%%%%%%%%%%%%%%
%%%%%%%%%%%%%%%%%%%%%%%%%%%%%%%%%%%%%%%%%%%%%%%%%%%%%%%%%%%%%%%%%%%%%%%%%%%%%%%%%%%%%%%%%%%%%%%

\begin{figure}[t]
	\includegraphics[width=3.4in]{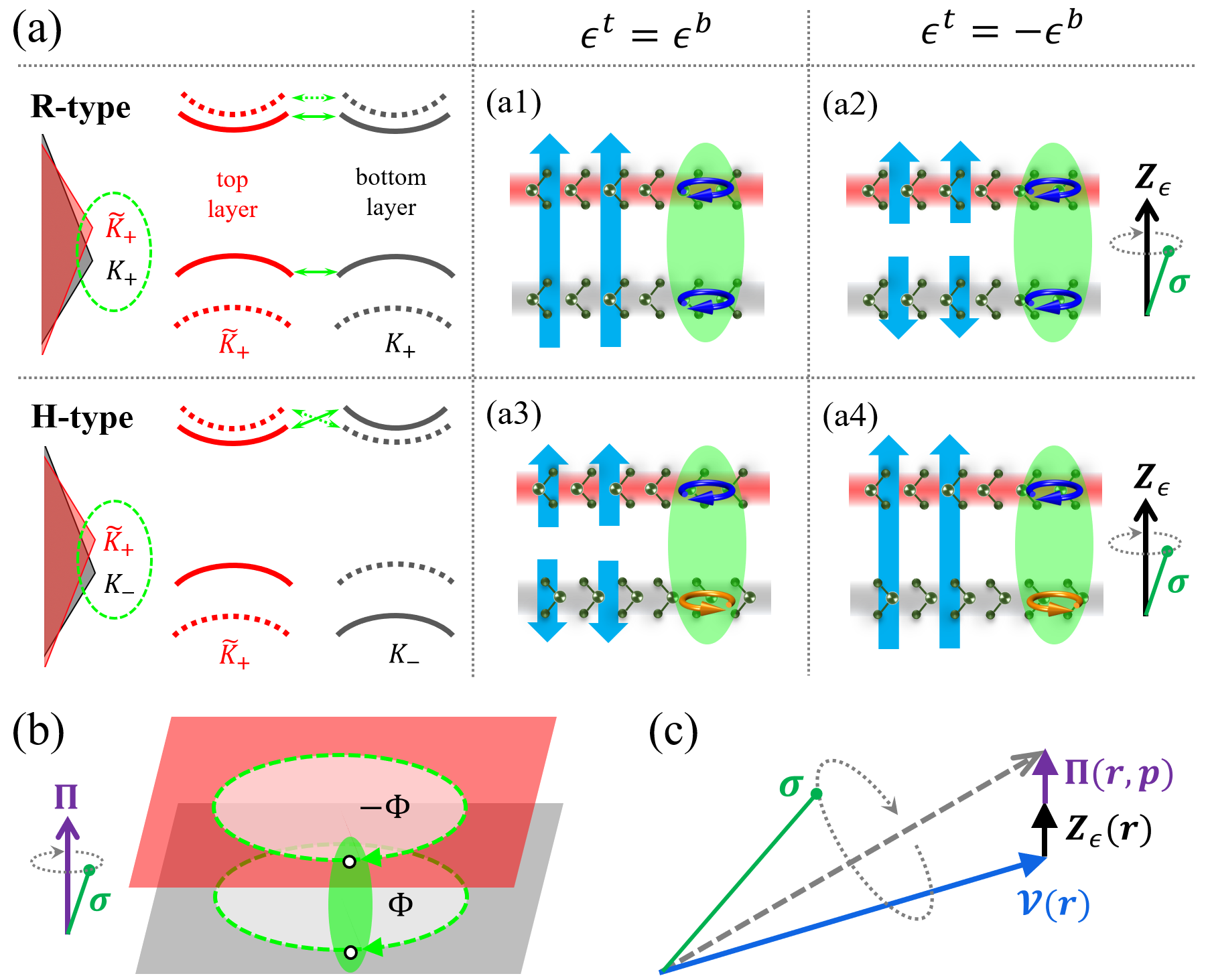}
	\caption{Comparison of twisted homobilayer TMD (e.g. MoX$_2$) with different stacking and strain configurations. (a) First column: Monolayer Brillouin corners for R-type (upper) and H-type (lower) twisted bilayer. The dashed circles highlight the two coupled Dirac cones. Solid and dotted curves represent spin-split Dirac cones. Double-head arrows denote interlayer coupling. (a1) \& (a3): Strain induced pseudomagnetic field (wide blue arrows) and valley magnetic moment (curly arrows) for layer symmetric strain  ($\epsilon^t=\epsilon^b$) in R-type (upper) and H-type (lower) bilayer, respectively. (a2) \& (a4): Same as before but with layer antisymmetric strain ($\epsilon^t=-\epsilon^b$). Green stick and black arrow illustrate layer pseudospin precession around the strain Zeeman field $\bZ_{\epsilon}$. (b) Pseudospin precession around $\bPi$ in the presence of layer contrasted pseudomagnetic field. The AB phases gained in the two layers are opposite ($\pm\Phi$) when a particle evolves in a loop. (c) Schematics showing the pseudospin precessing around the net effective Zeeman field (gray dashed arrow) in the presence of both strain and moir\'e coupling.}
	\label{Fig:RHComparisonSchematics}
\end{figure}

In a long wavelength moir\'e subject to a general strain pattern, the low energy carriers are described by the continuum Hamiltonian~\cite{LiangFuStrainMoire,ZhaiPRMaterials,MikitoKoshinoLatticeRelaxation2019}
\begin{equation}
\mathcal{H}=
\begin{pmatrix} v\bpi_t\cdot\bsigma'_{\tau^t}+\frac{E_g}{2}\sigma'_z+V^t&U\\
U^{\dagger}& v\bpi_b\cdot\bsigma'_{\tau^b}+\frac{E_g}{2}\sigma'_z+V^b
\end{pmatrix}
\end{equation}
with $\bpi_l=\bp+\tau^l\bA^l_{\epsilon}$, $l=t,b$ the layer index, and $\tau^{l}=\pm$ the valley index in layer $l$. The effect of strain (tensor $\epsilon_{ij}$) is accounted by the vector potential $\bA^{l}_{\epsilon}= \frac{\sqrt{3}\hbar\beta}{2a}(\epsilon^l_{xx}-\epsilon^l_{yy},\,-2\epsilon^l_{xy})$~\cite{SupplementalMaterial}, where $a$ is the lattice constant, and $\beta\approx2-3$~\cite{StrainedTMDShiangFangPRB2018}. %\textcolor{red}{Here we focus on low energy physics with leading order strain effects by considering moderate strain intensity. Discussions on strain modified band edge locations~\cite{StrainedTMDShiangFangPRB2018}, and higher order strain effects~\cite{SpinConnectionStrainedGraphene,RenormalizedFermiVelocityStrainedGraphene}, can be found in the Supplemental Materials.}
$\bsigma'_{\tau^{l}}=(\tau^{l}\sigma'_{x},\,\sigma'_{y})$ and $\sigma'_z$ are Pauli matrices spanned by the metal d orbitals at the conduction ($c$) and valence ($v$) band edges~\cite{DiXiaoTMDPRL2012}. $V^{l}=\text{diag}(\tilde{V}^l_c,\, \tilde{V}^l_v)$ and $U=\begin{pmatrix}
\tilde{U}_{cc}&\tilde{U}_{cv}\\\tilde{U}_{vc}&\tilde{U}_{vv}
\end{pmatrix}$ accounts for the interlayer coupling, which are functions of interlayer registry with location dependence~\cite{HongyiPseudoFieldMoire,ZhaiPRMaterials,WuMacDonaldPRL2019,InterlayerCouplingTMDPRB2017}.
Because $U$ couples only states of the same spin [Fig.~\ref{Fig:RHComparisonSchematics}(a)], we consider one spin species per valley at a time. 

%It should be noted that interlayer coupling in R-type and H-type bilayers are of different natures [Fig.~\ref{Fig:RHComparisonSchematics}(a)]. States with identical valley and spin indices are coupled for R-type, and both conduction and valence bands can participate. In contrast, states from different valleys are coupled for H-type, and it is only allowed in the conduction bands due to the large spin-split valence band gap.\cite{SpinSplitingBilayerTMDNatComm2013,TMDChemSocRev2015} Furthermore, inversion symmetry in H-type bilayers imposes $\mathcal{V}_z\equiv\delta_{c}/2$, where $\delta_{c}$ is the spin split gap in conduction bands. 

%In the framework of local approximation, the moir\'e coupling $\mathcal{U}$ is determined by the local registry. In twisted bilayer with strain, $\mathcal{U}$ depends on displacements due to both rigid rotation and strain.\cite{MikitoKoshinoLatticeRelaxation2019,LeonBalentsLatticeRelaxation2019,EnaldievRelaxationTheory,MikitoKoshinoLatticeRelaxation2017}

The large band gap $E_g$ allows one to perturbatively eliminate the valence bands~\cite{WuMacDonaldPRL2019,HongyiPseudoFieldMoire,ZhaiPRMaterials}, to reach a reduced Hamiltonian on the electron
\begin{equation}
H
=\frac{1}{2m}\left(\bp+\bCalA_{\epsilon}\right)^2
+\mathcal{U} +\mathcal{Z}_{\epsilon}.\label{Eq:HvBeforeTransformation}
\end{equation}
It has a $2\times2$ matrix form, spanned by the layer pseudospin $\bsigma$. $\bCalA_{\epsilon}=\text{diag}(\tau^t \bA^t_{\epsilon},\,\tau^b \bA^b_{\epsilon})$ is the matrix of strain induced vector potential. 
Hereafter, we focus on the $\tau^t=+$ valley of the top layer, which is coupled with the $\tau^b=+$ ($-$) valley of the lower layer in the parallel (antiparallel) stacking [Fig.~\ref{Fig:RHComparisonSchematics}(a)]. Results for the other valley can be obtained by time-reversal symmetry.
%Bilayer valley index $\tau=\text{diag}(1,\,1)$ or $\text{diag}(1,\,-1)$ for R-type or H-type bilayers, respectively [Fig.~\ref{Fig:RHComparisonSchematics}(a)]. 

The term $ \mathcal{U}=
\begin{pmatrix}
\tilde{V}^t_c&\tilde{U}_{cc}\\
\tilde{U}^{*}_{cc}&\tilde{V}^b_c
\end{pmatrix} $ is responsible for the moir\'e potential and layer hybridization of the carriers~\cite{WuMacDonaldPRL2019,HongyiPseudoFieldMoire,ZhaiPRMaterials,SupplementalMaterial}. 
We can write $\mathcal{U}=\mathcal{V}_0 \sigma_0+ \bsigma\cdot\bCalV$, where $\sigma_0$ is the identity matrix. 
%$\mathcal{U}$ mixes and causes energy shifts in the otherwise decoupled strained layers. Its local eigenvalues read $E_\pm=\mathcal{V}_0\pm\mathcal{V}$, where $\mathcal{V}=\left|\bCalV\right|$. 
$\bCalV$ is an effective field that causes the layer pseudospin precession [Fig.~\ref{Fig:StrainedMoireAndMoirePotential}(a)]. 
Figs.~\ref{Fig:StrainedMoireAndMoirePotential}(b--c) plot its magnitude and in-plane texture in a moir\'e unit cell showing the strong location dependence, in  R- and H-type TMD bilayer examples, respectively.

$\mathcal{Z}_{\epsilon}$ is the Zeeman energy of the valley magnetic moment in the strain induced pseudomagnetic field $\bCalB_{\epsilon}=\frac{1}{e}\nabla\times\bCalA_{\epsilon}$. In monolayers, Dirac electron of effective mass $m$ carries an intrinsic magnetic moment $\tau\mu_B^{*}=\tau\frac{e\hbar}{2m}$ in out-of-plane direction~\cite{ValleyMagneticMomentPRL2007}. Interestingly, out of the three contributions to the magnetic moment in TMDs~\cite{ValleyMagneticMomentNatPhys2015}, 
the pseudomagnetic field couples only to this lattice contribution associated with the Berry phase, but not those from spin and atomic orbitals.
We can write $\mathcal{Z}_{\epsilon}=\bmu\cdot\bCalB_{\epsilon}$, where valley magnetic moment $\bmu$ is listed in Table~\ref{Table}.

%\textcolor{red}{$\bmu=\hat{\bf z} \frac{e\hbar}{2m} \sigma_0$ in R-type bilayer with identical valley magnetic moment from each layer [Figs.~\ref{Fig:RHComparisonSchematics}(a1--a2) curly arrows], while in H-type bilayer $\bmu=\hat{\bf z} \frac{e\hbar}{2m} \sigma_z$ [Figs.~\ref{Fig:RHComparisonSchematics}(a3--a4)] because the antiparallel stacking couples opposite valleys}~\cite{SpinSplitingBilayerTMDNatComm2013}. 

\begin{table}[b]
	\caption{R-type \textit{vs} H-type bilayer TMDs with inhomogeneous strain. $ B^{\alpha}_{\epsilon} \hat{\bf z} = \frac{1}{e} \nabla\times \bA^{\alpha}_{\epsilon}$, $Z^{\alpha}_{\epsilon} = \mu_B^{*} B^{\alpha}_{\epsilon} $, $\alpha=a,s$.  }\label{Table}
	
	\begin{ruledtabular}
		\begin{tabular}{c|ccccccc}
			&  $ \bmu$  & $\bCalA_{\epsilon}^{s}$ & $\bCalA_{\epsilon}^{a}$ & $\bCalB_{\epsilon}^{s}$ & $\bCalB_{\epsilon}^{a}$ & $\mathcal{Z}_{\epsilon}^{s}$ & $\mathcal{Z}_{\epsilon}^{a}$\\ [0.02in]
			\hline
			R-type & $ \mu_B^{*} \hat{\bf z}  \sigma_0$ & $\bA^s_{\epsilon}\sigma_{0}$ & $\bA^a_{\epsilon}\sigma_{z}$ & $ B^s_{\epsilon} \hat{\bf z} \sigma_{0}$ & $ B^a_{\epsilon} \hat{\bf z} \sigma_{z}$ & $Z^s_{\epsilon}\sigma_{0}$ & $Z^a_{\epsilon}\sigma_{z}$	 \\ [0.04in]
			H-type & $  \mu_B^{*} \hat{\bf z} \sigma_z$ &$\bA^s_{\epsilon}\sigma_{z}$ & $\bA^a_{\epsilon}\sigma_{0}$ & $ B^s_{\epsilon} \hat{\bf z} \sigma_{z}$ & $ B^a_{\epsilon} \hat{\bf z} \sigma_{0}$ & $Z^s_{\epsilon}\sigma_{0}$ & $Z^a_{\epsilon}\sigma_{z}$ \\ [0.04in]
		\end{tabular}
	\end{ruledtabular}
\end{table}

%We decompose a general strain into the layer symmetric (s) and antisymmetric (a) parts, defining $\bA^s_{\epsilon} \equiv (\bA^t_{\epsilon}+\bA^b_{\epsilon})/2$ and $\bA^a_{\epsilon} \equiv (\bA^t_{\epsilon}-\bA^b_{\epsilon})/2$. 
Using $\bA^s_{\epsilon} \equiv (\bA^t_{\epsilon}+\bA^b_{\epsilon})/2$ and $\bA^a_{\epsilon} \equiv (\bA^t_{\epsilon}-\bA^b_{\epsilon})/2$ to quantify the layer symmetric (s) and antisymmetric (a) parts of a general strain in bilayer, one decomposes $\bCalA_{\epsilon}=\bCalA_{\epsilon}^{s}+\bCalA_{\epsilon}^{a}$, as well as $\bCalB_{\epsilon}$.
%$\bCalA_{\epsilon}= \bA^s_{\epsilon} \sigma_0 + \bA^a_{\epsilon} \sigma_z$ in the R-type bilayer, while \textcolor{red}{the Pauli matrices are swapped in H-type}: $\bCalA_{\epsilon}= \bA^s_{\epsilon} \sigma_z + \bA^a_{\epsilon} \sigma_0$. 
The Hamiltonian then reads $H=\frac{1}{2m}\left(\bp+\bCalA_{\epsilon}^{s}+\bCalA_{\epsilon}^{a}\right)^2+\mathcal{U}+\mathcal{Z}_{\epsilon}^{s}+\mathcal{Z}_{\epsilon}^{a}$. Table~\ref{Table} summarizes the pseudospin dependence of these physical quantities. $\sigma_{0}$ \textit{vs} $\sigma_{z}$ reflect the parallel \textit{vs} antiparallel stacking in R-type and H-type bilayers. $\sigma_z $ marks the layer contrasted sign of the quantities, which will affect pseudospin dynamics as discussed next.

Remarkably, the two strain components, combined with the strain dependent valley magnetic moment $\bmu$, lead to four distinct scenarios of the pseudospin dynamics.
First, in the Zeeman term, it is the antisymmetric strain that leads to the splitting of the layer pseudospin in either bilayer stacking [Table~\ref{Table} last column, and Figs.~\ref{Fig:RHComparisonSchematics}(a2) \& (a4)]. 
Moreover, the geometric phase in the centre-of-mass (COM) motion due to layer contrasted pseudomagnetic field [wide blue arrows pointing oppositely in Figs.~\ref{Fig:RHComparisonSchematics}(a2) \& (a3), i.e. R-type (H-type) bilayer under antisymmetric (symmetric) strain] is another cause of pseudospin precession. Upon closing a loop in such fields, electron picks up opposite geometric phases $e^{\mp ie\Phi/\hbar}$ in its two layer components, resulting in an in-plane rotation of pseudospin [Fig.~\ref{Fig:RHComparisonSchematics}(b)], which is the AB effect in a SU(2) gauge field first discussed by Wu and Yang~\cite{NonAbelianABEffectWuYang}.
%This occurs in R-type bilayer under antisymmetric strain [Fig.~\ref{Fig:RHComparisonSchematics}(a2)], and H-type under symmetric strain [Fig.~\ref{Fig:RHComparisonSchematics}(a3)].
%While $\bPi_{\epsilon}^{\alpha}$, $\bCalK_{\epsilon}$ and $\bZ^a_{\epsilon}$ are caused by strain and they depend on different symmetry parts of strain. $\bPi_{\epsilon}^{\alpha}\times\bsigma$ arises due to the $\sigma_{z}$ component of $\bCalA_{\epsilon}$, which exhibits \textit{apparent non-Abelian} properties.

The above effects on pseudospin can be explicitly seen from its Heisenberg equation of motion,
\begin{equation}
	\dot{\bsigma} = -\frac{i}{\hbar}[H, \bsigma] =\frac{2}{\hbar}\left(\bPi_{\epsilon}+\bCalK_{\epsilon}+\bZ^a_{\epsilon}+\bCalV\right)\times\bsigma.\label{Eq:PseudoSpinEquationOfMotion}
\end{equation}
$\bPi_{\epsilon}=\frac{1}{2m}(\bp\cdot\bA^{\alpha}_{\epsilon}+\bA^{\alpha}_{\epsilon}\cdot\bp)\hat{\bf z}$, $\alpha=a/s$ for R-/H-type, reflects a pseudospin precession accompanying the COM motion. The aforementioned AB effect [Fig.~\ref{Fig:RHComparisonSchematics}(b)] is a manifestation of this term. $\bZ^a_{\epsilon}=Z^a_{\epsilon}\hat{\bf z}$ is from the Zeeman splitting in the antisymmetric strain (last column of Table~\ref{Table}). 
%While $\bPi_{\epsilon}^{\alpha}$, $\bCalK_{\epsilon}$ and $\bZ^a_{\epsilon}$ are caused by strain and they depend on different symmetry parts of strain. $\bPi_{\epsilon}^{\alpha}\times\bsigma$ arises due to the $\sigma_{z}$ component of $\bCalA_{\epsilon}$, which exhibits \textit{apparent non-Abelian} properties.\cite{GoldmanNonAbelianReview,NonAbelianABEffectWuYang,NonAbelianABEffectHorvathy} For example, a particle evolving in the presence of such a gauge potential in a closed loop will in general gain different phases $e^{\mp ie\Phi/\hbar}$ in the two pseudo-spin components, where $\Phi$ is the flux of $\bB^\alpha_{\epsilon}$ in the enclosed area [Fig.~\ref{Fig:RHComparisonSchematics}(b)]. $\bZ^a_{\epsilon}\times\bsigma$ is related to the antisymmetric strain component, which causes energy splitting in the two layers [Fig.~\ref{Fig:RHComparisonSchematics}(c)]. Different potential depths in the two layers cause layer asymmetric charge distributions. 
$\bCalK_{\epsilon}=\frac{1}{m}\bA^s_{\epsilon}\cdot\bA^a_{\epsilon}\hat{\bf z}$ is a cross term between symmetric and antisymmetric strain components, which is negligible under modest strain. 
%Thus it is absent for pure layer symmetric/antisymmetric strain configurations in Fig.~\ref{Fig:RHComparisonSchematics}. 
$\bZ^a_{\epsilon}$ is function of location, while $\bPi_{\epsilon}$ also depends on momentum, both pointing out-of-plane. 
In a strain of $\epsilon \sim 1 \%$ and at Fermi wavelength $\lambda_f \sim 2$ nm (corresponding to Fermi energy of $\sim 20$ meV in TMDs), $\Pi_{\epsilon} \sim \frac{\hbar^2}{2 m a} \frac{\epsilon}{\lambda_F} \sim 1$ meV.
And for strain variation $\Delta \epsilon \sim 1 \%$ over a length $l \sim 10$ nm, $Z^a_{\epsilon} \sim \frac{\hbar^2}{2 m a} \frac{\Delta \epsilon}{ l} \sim 0.2$ meV. %$Z^a_{\epsilon} \sim \frac{\hbar^2}{2 m a} \frac{\Delta \epsilon}{ l}$, where $l$ is the length scale for strain to change by $\Delta \epsilon$If strain changes by $\Delta \epsilon \sim 1 \%$ over a length of $l \sim 10$nm, $0.2$meV, corresponding to a field of $2$Tesla.
%$\sim \frac{\hbar^2}{2 m a} \frac{\epsilon}{\lambda_F} \sim 1$ meV, if $\epsilon \sim 1 \%$ and $\lambda_f \sim 2$nm
%Fermi energy $\sim \frac{\hbar^2}{2 m } \frac{1}{\lambda_F^2} \sim 20$ meV, if $\lambda_f \sim 2$nm

In comparison, $\bCalV$ from the interlayer coupling has its orientation vary spatially [Figs.~\ref{Fig:StrainedMoireAndMoirePotential}(b--c)], determined by the local atomic registry in the moir\'e~\cite{ZhaiPRMaterials,HongyiPseudoFieldMoire,WuMacDonaldPRL2019}. Its magnitude $\mathcal{V} \sim $ O(1) meV at certain spots, and reaches O(10) meV over the rest area in the moir\'e, well exceeding that of $\bPi_{\epsilon}$ and $\bZ_{\epsilon}^a$. Fig.~\ref{Fig:RHComparisonSchematics}(c) illustrates the collective effect of these non-collinear pseudo-fields from both interlayer coupling and strain. As we show below, their interplay leads to genuine non-Abelian Berry phase effects that are absent with either moir\'e coupling or strain alone.

{\it Non-Abelian Berry curvature} - 
As the pseudospin dynamics is dominated by the interlayer coupling, it is natural to switch to the basis of its local eigenstates, satisfying $ \bsigma \cdot \bCalV \ket{\chi_{\pm} (\bf{r})}=\pm\mathcal{V}\ket{\chi_{\pm} (\bf{r})}$. 
%$\cos\zeta=\mathcal{V}_z/\mathcal{V}$ and $\sin\zeta=\left|\tilde{U}\right|/\mathcal{V}$, and $\phi$ is the azimuthal angle that equals the phase of $\tilde{U}^{*}$ [Fig.~\ref{Fig:StrainedMoireAndMoirePotential}(a)]. 
%In the following, $n_{\parallel}=\sin\zeta$ and $n_z=\cos\zeta$ will be used interchangeably. One can verify that $\braket{\bsigma}_{\pm}=\braket{\chi_{\pm}|\bsigma|\chi_{\pm}}=\pm\vec{n}$.
%Denote the eigenvector of $H$ as $\ket{\Psi}$, it reads $\ket{\Psi}=\sum_{i=\pm}\psi^i\ket{\chi_{i}}$ in the new basis, where $\psi^{i}$ characterizes the center-of-mass motion associated with the internal state $\ket{\chi_{i}}$. They are governed by the transformed Hamiltonian through $\tilde{H}\psi=E\psi$, where $\psi=\left(\psi^+,\,\psi^-\right)^T$.
The Hamiltonian in this basis reads 
\begin{equation}
\tilde{H}=Q^\dagger H Q=\frac{1}{2m}\left(\bp+\bCalA\right)^2 +E_{c} +\mathcal{\tilde{Z}}_{\epsilon}, \label{Eq:HvAfterTransformation}
\end{equation}
%$=\begin{pmatrix} \cos\zeta&\sin\zeta\\\sin\zeta&-\cos\zeta \end{pmatrix}$
where $Q=\left( |\chi_+ \rangle, \,| \chi_- \rangle \right)$.
The interlayer coupling is diagonalised, i.e. $E_c=\text{diag}(\mathcal{V}_0 + \mathcal{V},\, \mathcal{V}_0 - \mathcal{V})$, which characterises the scalar moir\'e potential experienced by the two pseudospin branches~\cite{HongyiPseudoFieldMoire}.
% is the moir\'e potential that specifies the spatially modulated band edge splittings. 
The gauge potential is now a composite one $\bCalA=\bCalA_{\nabla}+\boldsymbol{\mathcal{\tilde{A}}}_{\epsilon}$. The first part is due to the transformation in the non-Abelian group~\cite{QianNiuRMP}, 
$\bCalA_{\nabla}=-i\hbar Q^{\dagger}\nabla Q=-\frac{\hbar}{2}\left(\nabla\phi\right)\tilde{\sigma}_{z}
+\frac{\hbar}{2}\left(\nabla\zeta\right)\sigma_{y}$,
where $\zeta (\bf{r})$ and $\phi (\bf{r})$ are the polar and azimuthal angles of the local $\bCalV$ respectively [Fig.~\ref{Fig:StrainedMoireAndMoirePotential}(a)], and $\tilde{\sigma}_z \equiv Q^\dagger\sigma_{z}Q=\sigma_{z}\cos\zeta+\sigma_{x}\sin\zeta$. Its explicit form is gauge dependent and we adapt the one in Ref.~\cite{ZhaiPRMaterials}.
%obtained from Eq.~(\ref{Eq:QExpression}), 
%which exists even in the absence of strain.\cite{ZhaiPRMaterials}
The second part is from the strain, 
$\boldsymbol{\mathcal{\tilde{A}}}_{\epsilon}=Q^{\dagger}\bCalA_{\epsilon}Q
=\bA^{\alpha_1}_{\epsilon}\sigma_{0}+\bA^{\alpha_2}_{\epsilon}\tilde{\sigma}_{z}$,
where $\alpha_1=s$ ($a$) and $\alpha_2=a$ ($s$) for R-type (H-type) bilayers (see Table~\ref{Table}). 
%Because of the $\bCalA_{\nabla}$ part, $[\mathcal{A}_{x},\mathcal{A}_{y}]\ne0$. 
The associated gauge field reads~\cite{GoldmanNonAbelianReview,QianNiuRMP}
%$ \bCalB=\frac{1}{e}\nabla\times\bCalA+\frac{i}{e\hbar}[\mathcal{A}_{x},\mathcal{A}_{y}]$.
%Straightforward calculations using the above expressions yield
\begin{eqnarray}
\bCalB
= \frac{1}{e}\nabla\times\bCalA+\frac{i}{e\hbar}[\mathcal{A}_{x},\mathcal{A}_{y}] \hat{\bf z}
=(B^{\alpha_1}_{\epsilon}\sigma_{0}+B^{\alpha_2}_{\epsilon}\tilde{\sigma}_{z}) \hat{\bf z}.
\end{eqnarray}
It corresponds to a transformation of the strain induced pseudomagnetic field $\bCalB = Q^{\dagger}\bCalB_{\epsilon} Q$ (Table~\ref{Table}). The Zeeman coupling of the valley magnetic moment to $\bCalB$ appears as the last term in Eq.~(\ref{Eq:HvAfterTransformation}), $\mathcal{\tilde{Z}}_{\epsilon}=Z^s_{\epsilon}\sigma_{0}+Z^a_{\epsilon}\tilde{\sigma}_{z}$.

The non-Abelian nature of the gauge field is evidenced from its noncommutativity at different locations:
$\left[\bCalB(\br),\bCalB(\br')\right]
=2i\sigma_{y}\text{B}^{\alpha_2}_{\epsilon}(\br)\text{B}^{\alpha_2}_{\epsilon}(\br')\sin\left(\zeta(\br')-\zeta(\br)\right)$.
This is endowed by the fact that the interlayer hopping $\mathcal{V}_x \sigma_x +  \mathcal{V}_y\sigma_y$ varies spatially in the moir\'e, and does not commute with the strain induced $\bCalA_{\epsilon}$ with a $\sigma_{z}$ part.
%The interplay of strain and interlayer coupling $\tilde{U}$ can therefore be described in terms of a non-Abelian gauge potential and gauge field. 
We can compare with situations where only either of the above causes is present. In the limit $\epsilon=0$, the gauge potential $\bCalA=\bCalA_{\nabla}$ is non-Abelian, but the gauge field simply vanishes~\cite{ZhaiPRMaterials}. In the limit of decoupled layers, $\zeta(\br) =0 $, the gauge field reduces to $\bCalB_{\epsilon}(\br)$ whose forms at different locations always commute. 
%Non-commutativity in $\bCalB$ is because the transformation based on moir\'e interlayer coupling is local, rendering different forms for $\sigma_{z}$ at distinct locations. 
In aligned bilayers, the ratio of interlayer hopping to the band offset ($\mathcal{V}_z$) is spatially uniform rendering $\zeta(\br)$ constant, which also ensures that $\bCalB$ at different locations commute. Indeed, the moir\'e pattern combined with the inhomogeneous strain creates a unique scenario for the genuine non-Abelian gauge field to emerge. 

A force operator can also be defined to illustrate the effects of the non-Abelian Berry phase effect on the COM motion: $\boldsymbol{\mathcal{F}}=m\dot{\bv}=-\frac{i}{\hbar}[m\bv,\tilde{H}]=-\frac{e}{2}\left(\bv\times\bCalB-\bCalB\times\bv\right)-\nabla\left(E_c+\mathcal{\tilde{Z}}_{\epsilon}\right)-\frac{i}{\hbar}[\bCalA,E_c+\mathcal{\tilde{Z}}_{\epsilon}]$, where $\bv=-\frac{i}{\hbar}[\br,\tilde{H}]=\frac{1}{m}\left(\bp+\bCalA\right)$ is the velocity operator.
The first term is the magnetic force by the non-Abelian gauge field, while the rest two terms are reminiscent of electric force~\cite{GoldmanNonAbelianReview}.
For comparison, in an unstrained moir\'e superlattice, the force reads $\boldsymbol{\mathcal{F}}_{\epsilon=0}=-\nabla E_c-\frac{i}{\hbar}[\bCalA_{\nabla},E_c]$ without the magnetic part. 

%Therefore, apart from modifying the electric force, strain also introduces a Lorentz force due to presence of $\bCalB$. This non-Abelian Berry curvature is able to induce interesting noncommutative phenomena as will be discussed in the following.

%%%%%%%%%%%%%%%%%%%%%%%%%%%%%%%%%%%%%%%%%%%%%%%%%%%%%%%%%%%%%%%%%%%%%%%%%%%%%%%%%%%%%%%%%%%%%%%
%%%%%%%%%%%%%%%%%%%%%%%%%%%%%%%%%%%%%%%%%%%%%%%%%%%%%%%%%%%%%%%%%%%%%%%%%%%%%%%%%%%%%%%%%%%%%%%

\begin{figure}[t]
	\includegraphics[width=3.4in]{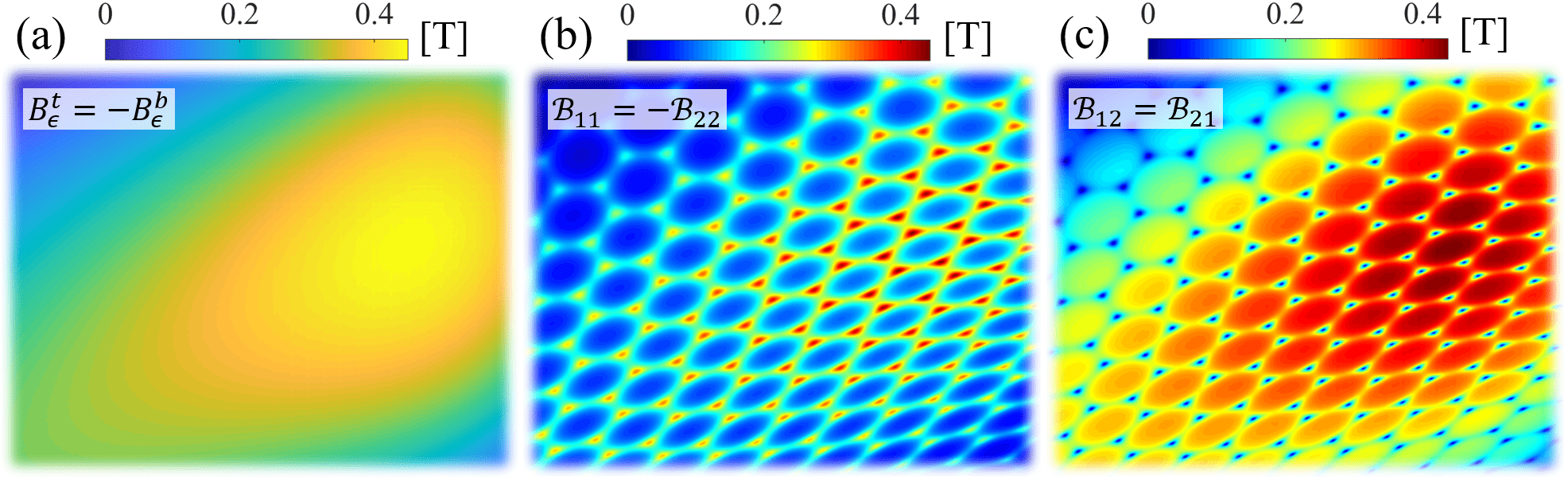}
	\caption{Strain pseudomagnetic field and non-Abelian gauge field in strained H-type $0.86^\circ$ twisted bilayer MoS$_2$. (a) $B^t_{\epsilon}=-B^b_{\epsilon}$. (b) $\mathcal{B}_{11}=-\mathcal{B}_{22}$. (c) $\mathcal{B}_{12}=\mathcal{B}_{21}$.}
	\label{Fig:BstrainAndBnA_expt}
\end{figure}

Fig.~\ref{Fig:BstrainAndBnA_expt} gives an example of the non-Abelian gauge field for H-type bilayer MoS$_2$ featuring the moir\'e pattern shown in Fig.~\ref{Fig:StrainedMoireAndMoirePotential}(c). The non-periodic moir\'e is produced by a $0.86^\circ$ twisting and a modest layer antisymmetric strain with peak magnitude $\epsilon^{a} \sim 0.5 \%$~\cite{SupplementalMaterial}. Oftentimes, inhomogeneous heterostrain of this magnitude is unintentionally introduced in the fabrication of moir\'e lattices~\cite{YusongStrainInTMDmoire}. Interestingly, such system naturally provides a platform for studying noncommutative phenomena.
%\cite{NonAbelianABEffectApplPhysB,NonAbelianABEffectScience2019,NonAbelianABEffectNatCommun2019,GoldmanNonAbelianReview}
%To illustrate the non-Abelian effects, we need to specify the details of strain. Intrinsic layer asymmetric strain are ubiquitous in experimentally available samples.\cite{BubblesWrinklesInTwistedBilayersNanoLett2014,YusongStrainInTMDmoire,FalkoRelaxationExpt,ImagingMoirePhilipKim,RelaxationTwistedBilayerNatMater2019} We first adapt a layer antisymmetric strain configuration observed in Ref.~\cite{YusongStrainInTMDmoire} and incorporate it into a H-type twisted bilayer MoS$_2$ [Fig.~\ref{Fig:StrainedMoireAndMoirePotential}(c--d)]. 
%Fig.~\ref{Fig:BstrainAndBnA_expt}(a) shows the distribution of the strain pseudo-magnetic field $B_{\epsilon}^t=-B_{\epsilon}^b$ (also see Fig.~\ref{Fig:BstrainAndBnA_expt_Htype_extended_area}). The field is rather weak due to modest strain. Figs.~\ref{Fig:BstrainAndBnA_expt}(b--c) present the resultant non-Abelian Berry curvature $\bCalB=\bB^t_{\epsilon}\tilde{\sigma}_z$ due to conduction band interlayer coupling. Its diagonal (off-diagonal) elements originate from the overlap of $B^t_{\epsilon}$ and out-of-plane (in-plane) component of $\hat{\bn}$ (Fig.~\ref{Fig:n_expt}). In this case, $\hat{\bn}$ points in the upper half plane, hence $\bCalB$ is non-negative.

\begin{figure}[t]
	\includegraphics[width=3.4in]{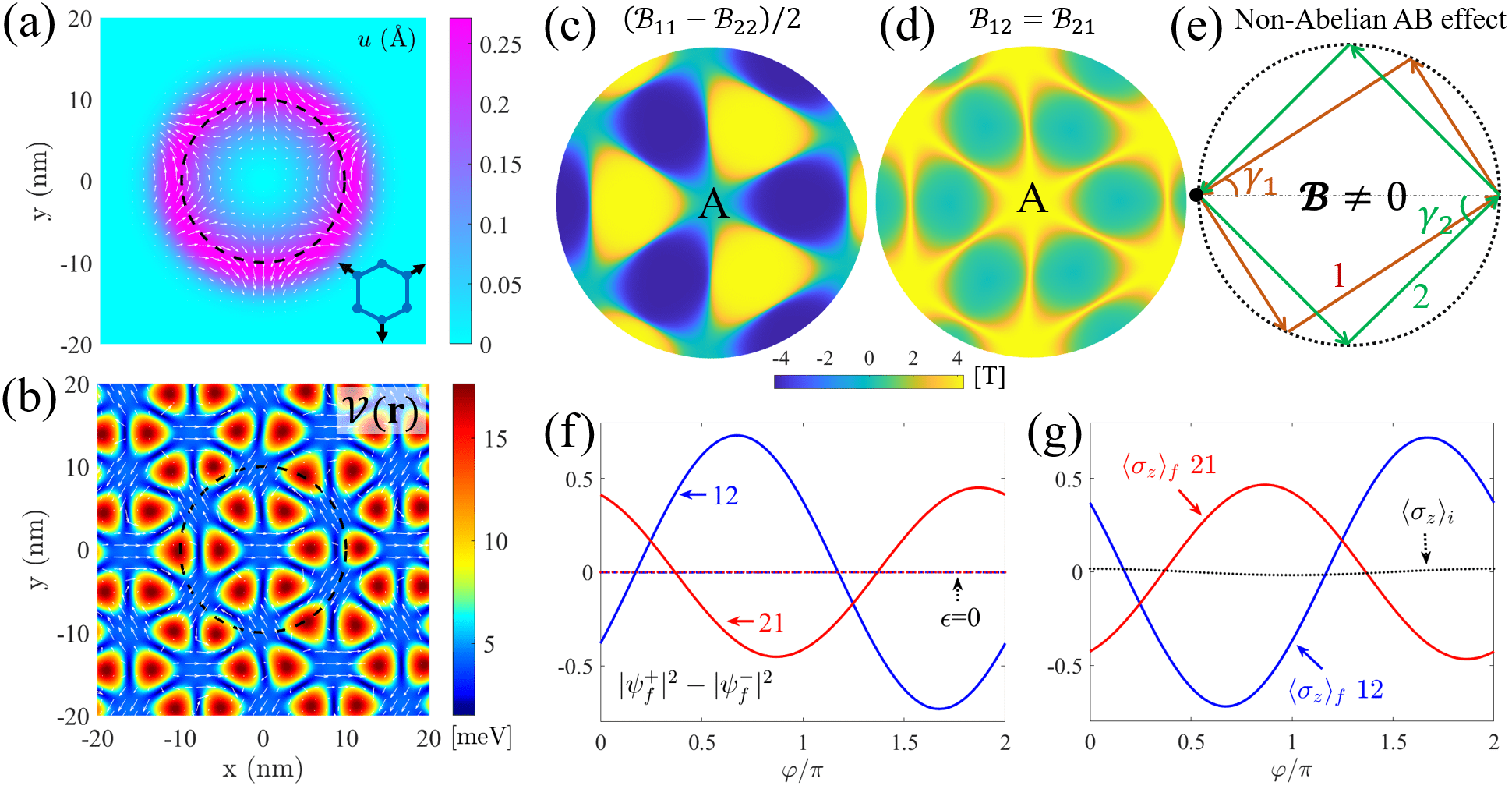}
	\caption{Non-Abelian gauge field and AB effect in triaxially strained R-type $2^\circ$ twisted bilayer MoSe$_2$. (a) Magnitude (background) and vector flow (arrows) of the strain displacement. Inset shows schematics of triaxial stress. (b) Distribution of $\mathcal{V}$ with arrows denoting the in-plane component of $\hat{\bCalV}$. (c) Diagonal and (d) off-diagonal components of the non-Abelian gauge field. Center of the strained area has the A stacking. (e) Schematics of non-Abelian AB effect setup with loop ordering $1\rightarrow2$ or $2\rightarrow1$. $\gamma_{1}=30^\circ$ and $\gamma_{2}=40^\circ$. (f) Population redistribution in the two internal states $\ket{\chi_\pm}$ \textit{vs} $\varphi$ when the particle travels $1\rightarrow2$ (solid blue) or $2\rightarrow1$ (dashed red). (g) Population redistribution in the two layers $\braket{\sigma_{z}}$ at the starting/ending point [black dot in (e)] \textit{vs} $\varphi$ after the particle travels the aforementioned paths. Black dotted curve shows $\braket{\sigma_{z}}$ for the initial state.}
	\label{Fig:BnA_triaxial}
\end{figure}

%Next we consider the valence band coupling in a R-type twisted homobilayer MoSe$_2$ with externally engineered strain. In contrast to the intrinsically strained sample considered above, 
The moir\'e lattice and strain can also be independently controlled to engineer the non-Abelian gauge field. 
%For simplicity, we assumed that only the bottom layer is triaxially strained (see SM and Fig.~\ref{Fig:Displacement_and_Strain_tensor_triaxial} for more details). In experiments, triaxial strain can be introduced by directly stretching the sample,\cite{TriaxialStrainFromForce,TriaxialStrainNatPhys,StrainFromForceNanoLett} or depositing the sample onto a triangular substrate.\cite{TriaxialStrainSciAdv2019,YuhangNanoLett2017}
Stress of various forms can be applied first on the bottom layer using movable substrates~\cite{TriaxialStrainFromForce}, before the top layer is deposited with the control of twisting angle.
Strain can also be introduced by depositing 2D materials on pre-engineered substrates with tailorable shapes and sizes~\cite{TriaxialStrainSciAdv2019,YuhangNanoLett2017}.
Fig.~\ref{Fig:BnA_triaxial} illustrates an example of R-type bilayer MoSe$_2$, whose bottom layer is triaxially strained over a circular area~\cite{SupplementalMaterial}. Fig.~\ref{Fig:BnA_triaxial}(b) plots the interlayer coupling $\bCalV (\bf{r})$ in the strain distorted moir\'e landscape under a $2^\circ$ twist of top layer, and a peak strain magnitude $\epsilon^{b} \sim 0.5 \%$. 
%where $u_0$ is a constant determining the maximum intensity of the strain.\cite{TriaxialStrainPeeters} The strain tensor can be estimated as $\epsilon=\begin{pmatrix}2y&2x\\2x&-2y\end{pmatrix}$. As the strained area is usually finite in experiments, the displacement and strain should relax after certain distance. 
Figs.~\ref{Fig:BnA_triaxial}(c-d) show the non-Abelian gauge field $\bCalB$ within the strained area, which can reach a few Tesla. Intensity and landscape of the non-Abelian gauge field can be tuned via strain, twist angle, as well as interlayer bias~\cite{SupplementalMaterial}.
%In order to have pronounced non-Abelian features, it is desirable to have small gaps between $E_{\pm}$ and rapid spatial variations in the orientation of $\vec{n}$. Otherwise adiabatic approximation might be applicable and Abelian features dominate at low energy.\cite{HongyiPseudoFieldMoire, ZhaiPRMaterials,NonAbelianGaugeRMP,GoldmanNonAbelianReview} An interlayer bias can be applied to reduce the spacing between $E_{\pm}$, and twist angle can be enlarged to increase spatial variation of $\vec{n}$ (see SM for more details).\cite{HongyiPseudoFieldMoire,ZhaiPRMaterials} 

{\it Genuine non-Abelian AB effect} - 
%For the observation of physical effects from the non-Abelian Berry curvature, we propose an AB effect-like scheme in the high energy regime. When the kinetic energy of the particle is much higher than the scalar potentials, one can only keep the non-Abelian gauge potential in the Hamiltonian, i.e. $-\frac{1}{2m}\left(\bp+\bCalA\right)^2\psi\approx E\psi$.\cite{GeometricPhasePRL2020} When a particle enters the strained region, it will gain a phase $\mathcal{P}=P\text{exp}\left(-\frac{i}{\hbar}\int_{\br_0}^{\br'}\bCalA\cdot d\bl\right)$ after traveling from $\br_0$ to $\br'$, where $P$ represents path-ordering. 
Evolution in a SU(2) gauge potential can generally lead to the change of particle's pseudospin due to the geometric phase  $\mathcal{P}=P\text{exp}\left(-\frac{i}{\hbar}\int_{\br_0}^{\br'}\bCalA\cdot d\bl\right)$, where $P$ denotes path-ordering. 
However, such evolution is not necessarily non-Abelian. For example, the strain induced gauge field in decoupled bilayer can only lead to pseudospin precession about the $z$ direction which is Abelian [Fig. 2(b)]. In the distorted moir\'e with both heterostrain and interlayer coupling, the noncommuting nature of the gauge field underlies a genuine non-Abelian evolution~\cite{GoldmanNonAbelianReview}, which can be illustrated by the following AB effect.

Initially on pseudospin $\ket{\psi_i}=  \cos\eta \ket{\chi_+} + e^{i\varphi}\sin\eta \ket{\chi_-}$, we compare the final state $\ket{\psi_f} = \mathcal{P} \ket{\psi_i} = \psi_f^+ \ket{\chi_+} + \psi_f^- \ket{\chi_-}$ after the electron travels two closed loops in the strained area in different orders [$1\rightarrow2$ or $2\rightarrow1$, Fig.~\ref{Fig:BnA_triaxial}(e)].
%Specifically, one expects $\left|\psi^{+}_f\right|^2-\left|\psi^{-}_f\right|^2\ne\left|\psi^{+}_i\right|^2-\left|\psi^{-}_i\right|^2$, where $\psi_i=\left(\psi_i^+,\,\psi_i^-\right)^T$ and $\psi_f=\left(\psi_f^+,\,\psi_f^-\right)^T=\mathcal{P}\psi_i$ denote the incoming and outgoing states.\cite{NonAbelianABEffectApplPhysB,NonAbelianABEffectScience2019,NonAbelianABEffectNatCommun2019,GoldmanNonAbelianReview} Furthermore, such population redistribution between the two internal states is expected to be path-dependent if a non-Abelian Berry curvature exists.\cite{NonAbelianABEffectApplPhysB,NonAbelianGaugeRMP} Outside the strained region, the initial state can be prepared as $\psi_i=\left(\cos\eta,\,e^{i\varphi}\sin\eta\right)^T$.\cite{NonAbelianABEffectApplPhysB,NonAbelianABEffectScience2019} 
%For illustration purposes, we select paths formed by combining two right triangles inscribed in the circle [Fig.~\ref{Fig:BnA_triaxial}(c)], which can be parameterized by angles $\gamma_{1,2}$, respectively. Fig.~\ref{Fig:BnA_triaxial}(d) shows the distinct population redistribution as function of $\varphi$ when the particle follows the counterclockwise (blue) or clockwise (red) path, where $\gamma_{1}=\gamma_{2}=\pi/6$ is employed, and 
Fig.~\ref{Fig:BnA_triaxial}(f) plots $|\psi_f^+|^2 - |\psi_f^-|^2$, as a function of the initial state phase angle $\varphi$, with $\eta=\pi/4$. The discrepancy between the two differently ordered paths is clearly seen, a signature of the genuine non-Abelian AB effect~\cite{GoldmanNonAbelianReview}. Note that dynamical phase can also lead to a pseudospin precession, determined by the path integral of $\mathcal{\tilde{Z}_{\epsilon}}(\br)$ and $\mathcal{V}(\br)$. This, however, is expected to be loop-order independent and does not affect $|\psi_f^+|^2 - |\psi_f^-|^2$.%\footnote{It is clearer by using the representation mentioned in Ref.~\cite{CommentsOnAlternativeTransformation} that the dynamical phase is independent of loop order as $\mathcal{\tilde{Z}_{\epsilon}}$ is absent. In order to reduce the effect of the dynamical phase, one might employ modest strain and reduce the strength of interlayer coupling (e.g. by adding hBN spacer between the bilayers~\cite{hBNspacer}) so that kinetic energy dominates~\cite{GeometricPhasePRL2020}.}.

We note that the $z$ component of the pseudospin corresponds to a measurable quantity, i.e. out-of-plane electrical polarization. 
%Denote its value evaluated with initial and final state at $\br_0$ and $\br'$ as $\braket{\sigma_z}_{i}$ and $\braket{\sigma_z}_{f}$, respectively.
%In general, one expects that $\braket{\sigma_z}_{f}\ne\braket{\sigma_z}_{i}$ and $\braket{\sigma_z}_{f}$ is path dependent if $\bCalB\ne0$. 
Red and blue curves in Fig.~\ref{Fig:BnA_triaxial}(g) give the corresponding plots of the final state polarization $\braket{\sigma_z}$ for the evolutions in Fig.~\ref{Fig:BnA_triaxial}(f). The difference between the two differently ordered paths is also seen, both of which are distinct from the initial value (black dotted curve).

For comparison, we examine evolutions on the same pair of loops with different orders in the unstrained moir\'e, where the gauge field $\bCalB$ vanishes. These are shown by the dashed lines in Fig.~\ref{Fig:BnA_triaxial}(f). Although the SU(2) Berry connection $\bCalA_{\nabla}$ is finite and non-Abelian, it does not have any effect on $|\psi_f^+|^2 - |\psi_f^-|^2$. Electrical polarization $\braket{\sigma_z}$ of the final states is also found to be identical to that of initial ones. 
%the dashed lines are for the case without strain, where the Berry curvature $\bCalB$ vanishes. Having the non-Abelian Berry connection $\bCalA_{\nabla}$ only, $|\psi_f^+|^2 - |\psi_f^-|^2$ is not affected. 
Likewise, in decoupled bilayer, the strain induced pseudomagnetic field alone does not change $\braket{\sigma_z}$. In either scenarios, one can not distinguish the two loop orderings, $1\rightarrow2$ \textit{vs} $2\rightarrow1$.
The genuine non-Abelian AB effect signifies the profound role of the non-Abelian gauge field $\bCalB$, which arises from the interplay of inhomogeneous strain and interlayer coupling in the distorted moir\'e only.

%The conventional AB effect proves the physically measurable significance of the gauge potential. Its non-Abelian version further shows the noncommutativity in non-Abelian gauge theory. Fig.~\ref{Fig:BnA_triaxial} illustrates an idealized scenario, its realization is technically challenging that relies on advances in nanoelectronics for better manipulation of electron trajectory and readout of final pseudospin states. On the other hand, non-Abelian gauge field may manifest in other transport phenomena. For instance, moir\'e lattices enable experimental exploration of the Hofstadter butterfly spectrum. Non-Abelian gauge fields are expected to generate distinct fractal patterns compared to their Abelian counterparts~\cite{NonAbelianABEffectScience2019}. Such phenomena deserves further exploration, and whether integer quantum Hall effect is still observable in the non-Abelian regime also remains unresolved. Strained moir\'e lattices with the emergent non-Abelian gauge field provide an electronic system to study such quantum Hall physics. Moreover, the 2D nature and intriguing interlayer coupling in moir\'e latices enable strong Coulomb interaction. Thus, manifestation of noncommutativity of non-Abelian gauge field in the quantum many-body phenomena~\cite{GoldmanNonAbelianReview} is also an interesting future research direction.

While the conventional AB effect proves the physically measurable significance of gauge potential, its genuine non-Abelian version can be as fundamental in demonstrating the noncommutativity in non-Abelian gauge theory. Realization of the scenario in Fig.~\ref{Fig:BnA_triaxial}(e) relies on advances in nanoelectronics for manipulation of electron trajectory and readout of final pseudospin states. On the other hand, non-Abelian gauge field can have interesting manifestations in other transport phenomena. For instance, non-Abelian gauge fields are expected to generate distinct fractal patterns in the Hofstadter butterfly spectrum compared to their Abelian counterparts~\cite{NonAbelianABEffectScience2019}, and whether integer quantum Hall effect is still observable in the non-Abelian regime also remains unresolved. Strained moir\'e superlattices can be an ideal arena for their exporations. Moreover, moir\'e superlattice has also proven to be a powerful experimental platform to explore quantum many-body phenomena, where the manifestation of noncommutativity of non-Abelian gauge field is also highly interesting~\cite{GoldmanNonAbelianReview}.

%\textcolor{blue}{(Math details, comment out later) $\braket{\sigma_z}(\br)=\braket{\Psi|\sigma_{z}|\Psi}=\left(\cos2\eta\right)\cos\zeta+\left(\sin2\eta\right)\left(\cos\varphi\right)\sin\zeta$.
%\begin{equation}
%\begin{aligned}
%\braket{\sigma_z}(\br)
%&=\braket{\Psi|\sigma_{z}|\Psi}\\
%&=\left(\left|\psi^{+}\right|^2-\left|\psi^{-}\right|^2\right)\cos\zeta+2\text{Re}\left(\psi^{+}\psi^{-*}\right)\sin\zeta\\
%&=\left(\cos2\eta\right)\cos\zeta+\left(\sin2\eta\right)\left(\cos\varphi\right)\sin\zeta.
%\end{aligned}	
%\end{equation}
%For the initial state characterized by $\eta=\pi/4$ and $\varphi_0$, one has $\braket{\sigma_z}_{i}=\left(\cos\varphi_0\right)\sin\zeta_0$, where $\zeta_0$ is the value of $\zeta$ at the starting point $\br_0$. Assume that final state is characterized by $\eta'$ and $\varphi'$ after the evolution, its corresponding layer pseudo-spin reads $\braket{\sigma_z}_{f}=\left(\cos2\eta'\right)\cos\zeta'+\left(\sin2\eta'\right)\left(\cos\varphi'\right)\sin\zeta'$, where $\zeta'$ denotes the value of $\zeta$ at the final location $\br'$.}

%%%%%%%%%%%%%%%%%%%%%%%%%%%%%%%%%%%%%%%%%%%%%%%%%%%%%%%%%%%%%%%%%%%%%%%%%%%%%%%%%%%%%%%%%%%%%%%
%%%%%%%%%%%%%%%%%%%%%%%%%%%%%%%%%%%%%%%%%%%%%%%%%%%%%%%%%%%%%%%%%%%%%%%%%%%%%%%%%%%%%%%%%%%%%%%
\section{Acknowledgment}
We thank Yusong Bai for providing the data of strained moir\'e structure and Qizhong Zhu for helpful discussions. The work is supported by the Research Grants Council of Hong Kong (Grants No. HKU17306819 and No. C7036-17W), and the University of Hong Kong (Seed Funding for Strategic Interdisciplinary Research).

%===========================================================================================
\bibliography{Refs}
\bibliographystyle{apsrev4-1}

\end{document}